\documentclass[preprint,twocolumns,times]{elsarticle}
\usepackage{amsmath}
\usepackage{graphicx}
\usepackage{color}
\usepackage{CJK}
\usepackage{dcolumn}   % needed for some tables
\usepackage{hyperref}
\biboptions{numbers,sort&compress}
\usepackage{setspace}

\begin{document}
	
	\begin{frontmatter}
		
		\title{Nonlinear interaction of head$-$on solitary waves in integrable and nonintegrable systems}
		
		\author{Shutian Zhang} \author{Shikun Liu} \author{Tengfei Jiao} \author{Min Sun}
		\author{Decai Huang \corref{Huang}}
		
		\cortext[Huang]{hdc@njust.edu.cn, 086-15951864599}
		%\author[NUST]{Decai Huang \corref{Huang}}
		\address{Department of Applied Physics, Nanjing University of Science and Technology, Nanjing 210094, China}%[NUST]
		
		\date{\today}
		
		\begin{abstract}
			
			This study numerically investigates the nonlinear interaction of head$-$on solitary waves in a granular chain (a nonintegrable system) and compares the simulation results with the theoretical results in fluid (an integrable system). Three stages (i.e., pre$-$in$-$phase traveling stage, central$-$collision stage, and post$-$in$-$phase traveling stage) are identified to describe the nonlinear interaction processes in the granular chain. The nonlinear scattering effect occurs in the central$-$collision stage, which decreases the amplitude of incident solitary waves. Compared with the leading$-$time phase in the incident and separation collision processes, the lagging$-$time phase in the separation collision process is smaller. This asymmetrical nonlinear collision results in an occurrence of leading phase shifts of time and space in the post$-$in$-$phase traveling stage. We next find that solitary wave amplitude does not influence the immediate space$-$phase shift in the granular chain. The space$-$phase shift of the post$-$in$-$phase traveling stage is only determined by measurement position rather than wave amplitude. The results are reversed in the fluid. An increase in solitary wave amplitude leads to decreased attachment, detachment and residence times for granular chain and fluid. For the immediate time$-$phase shift, leading and lagging phenomena appear in the granular chain and the fluid, respectively. These results offer new knowledge for designing mechanical metamaterials and energy$-$mitigating systems.
			
		\end{abstract}
		
		
	\end{frontmatter}
	
	\section{Introduction}
	\label{Intro}
	
	Nonlinear waves have been observed extensively in nature, such as in oceans, plasmas, solid lattices, biological molecules and optical systems \cite{Osborne1980SCI208.451,Shukla1984POF27.327,Shukla2007PLA367.120,Schwarz1999PRL83.223,Xie2000PRL84.5435,Fleischer2003Nature422.147}. A solitary wave (SW) is unique because of its localization and stabilization properties \cite{Wazwaz2009,Sen2008PR462.21,Rosas2018PhysRep753.1}. The classical Korteweg$-$de Vries (KdV) equation, which is integrable, was first introduced to describe the traveling dynamics of SW in shallow water \cite{Korteweg1895PM39.422}. The other known integrable equations have analytical SW solutions, such as the sine$-$Gordon equation and the nonlinear Schr$\ddot{\rm o}$dinger equation \cite{Sen2008PR462.21}. However, many nonlinear systems admitting solitary waves (SWs), such as granular chain (GC) and the Fermi$-$Pasta$-$Ulam$-$Tsingou system, are nonintegrable \cite{Nesterenko1983JAMTP24.733,Nesterenko2018PTRSA376.20170130,Fermi1955,Vainchtein2022PhysD434.133252}. In general, a SW with larger amplitude has a higher wave speed for integrable and nonintegrable systems. For the former, increasing the amplitude of SWs can decrease the width of SW \cite{Sen2008PR462.21}. For the latter, Nesterenko first found the existence of SW in a sonic$-$vacuum GC whose width is independent of the wave amplitude \cite{Nesterenko2001}. Following his pioneered work, massive activities involving SWs in GC are focused on the excitation of SW, the reflection and transmission of SW at an interface, and the interaction of SWs \cite{Sen1998PRE57.2386,Sen1999PhysA268.644,Manciu1999PhysA274.588,Chong2017JPConMat29.413003,Manciu2000PRE63.016614,Manciu2002PRE66.016616,Avalos2009PRE79.046607,Santibanez2011PRE84.026604,Shen2014PRE90.022905,Wang2018CPB27.044501,Wu2020CPL37.074501,Zhang2021IJMS191.106073,Tichler2013PRL111.048001,Liu2015PRE92.013202,Yang2016AIP6.075317,Du2020IJP98.1249}. Therefore, comparing the traveling dynamics between integrable and nonintegrable systems is of great significance.
	
	Given the intrinsic characteristic of the interaction between the touching grains in GC, such as Hertzian force for spherical grain, the key difference is reflected in the interactions between the SWs compared with those in integrable systems. Experimental and numerical studies play a great role because of the nonintegrability of the constitutive equations \cite{Sen1999PhysA268.644,Manciu1999PhysA274.588,Chong2017JPConMat29.413003}. To create an SW in GC, a commonly used method is an instantaneous impulse imparted on the sidemost grain. Sen et al. found that the central grain remains static in an odd chain, and the oscillated motions of the grains around the center are observed, which leads to the formation of secondary SWs \cite{Manciu2000PRE63.016614,Manciu2002PRE66.016616,Avalos2009PRE79.046607}. The occurrence of secondary SWs was convinced by experimental results \cite{Santibanez2011PRE84.026604}. To compare the difference of the collision of SWs in integrable and nonintegrable systems, Shen et al. directly introduced the analytical solutions of SW of the KdV equation and Toda lattice into the GC \cite{Shen2014PRE90.022905}. In the integrable systems, the nonlinear collision effect is observed, which leads to a phase shift appearing though the SWs after the collision recovers the amplitude and propagation velocity of the incident SWs. The phase shift is also observed in a nonintegrable GC system owing to a nonlinear collision. However, the nonlinear scattering effect occurs, leading to the generation of secondary SWs, and the amplitude of SWs decreases after the collision. Whether to generate the phase shift is still a controversial issue. Wang et al. employed the KdV solution as the incident waves to study collision dynamics of head$-$on propagating solitary waves (HSWs) \cite{Wang2018CPB27.044501}. The simulation results indicate that the collision of the HSWs does not influence the waveform and amplitude and no phase shift happens. In our previous study, the phase shift arises after HSWs collided in GC when the analytical SW solution of the Toda lattice was used as incident SWs \cite{Wu2020CPL37.074501}. Further results show that the collision of HSWs can be divided into two processes: early incident collision and latter separation collision processes. Lagging and leading phases are discovered respectively. The latter's effect is stronger than the former, resulting in a leading phase after the collision.
	
	The collision dynamics of HSWs is also a critical topic in integrable systems \cite{Su1980JFM98.509,Fenton1982JFM118.411,Maxworthy1976JFM76.177,Chambarel2009NonPG16.111,Chen2014JFM749.577,Tong2019WM88.34,Deng2020Chaos30.043101}. Taking continuous fluid for example, Su et al. carried out a third$-$order perturbation analysis on the collision of HSWs \cite{Su1980JFM98.509}. They found that an increase in the amplitude of HSWs increases the immediate space$-$phase shift (at the collision center) and the uniform space$-$phase shift (far away from the collision center). When the HSWs are two identical SWs, the collision dynamics of HSWs is found equivalent to that of a single SW and a vertical wall \cite{Maxworthy1976JFM76.177,Fenton1982JFM118.411,Power1984WM6.183,Byatt-Smith1988JFM197.503,Cooker1997JFM342.141,Chen2015EJMBF49.20}. Cooker et al. explored the time$-$phase shift by introducing the definitions of the attachment, detachment, and residence times \cite{Cooker1997JFM342.141}. The results show that these characteristic times decrease with the increased amplitude of HSWs. Similar results were reproduced in integrable plasma systems \cite{ZhangJ2014POP21.103706,Qi2014POP21.082118,ZhangJ2016CPL33.065202}. For the collision of two identical HSWs, analytical and simulated results evidenced that increasing wave amplitude can increase space$-$phase shift. The reported results have shown that the collision dynamics of HSWs share a common characteristic, that is, phase shift, in integrable and nonintegrable systems. The SWs after the collision can recover their initial incident waveform for the former. For the latter, the scattering effect due to the nonlinear collision leads to secondary SWs, and the amplitude of SWs decreases after the collision. Therefore, borrowing the theoretical analysis methods used in integrable systems to describe the collision dynamics of HSWs in nonintegrable GC is of great significance.
	
	This study explored the nonlinear interaction of two identical HSWs in a nonintegrable sonic$-$vacuum GC. The complex collision process is elucidated using the theoretical analysis method in fluid. The structure of this study is as follows: Section \ref{SimuMod} sets up the simulation model for an unprecompressed GC, Section \ref{TheoAna} revisits the theoretical analysis of the collision of HSWs in GC and fluid, Section \ref{SimuResDis} presents the simulation results and compares them with those in fluid (i.e., space$-$phase shift and time$-$phase shift), Section \ref{Conc} draws our conclusions.
	
	\section{Simulation model}
	\label{SimuMod}
	
	The discrete element method is used to explore the collision characteristic of HSWs in a one$-$dimensional monodisperse GC comprised of $N$ spherical elastic beads with mass $m$ and radius $R$. The interaction between two neighboring beads, without dissipation, is modeled by Hertz potential \cite{Landau1959}.
	\begin{equation}
		\begin{split}
			U=\left\{
			\begin{array}{ll}	
				\frac{2}{5}k\delta^{5/2}, & \delta\geq0\\
				0, & \delta<0
			\end{array},
			\right.
		\end{split}
		\label{eq:Hertz potential}
	\end{equation}
	where $\delta$ is the overlap deformation between touching beads and $k={E\sqrt{2R}}/{3(1-\sigma^2)}$ is the elastic coefficient which reflects on the material and geometric properties of the beads. $E$ and $\sigma$ are Young's modulus and Poisson ratio, respectively. In the simulation, an unprecompressed GC is considered, which is always called sonic$-$vacuum GC. For the touching beads, the equation of motion for bead $i$ is given by
	\begin{equation}
		m\ddot{s_{i}}=k\left[\left(s_{i-1}-s_{i}\right)^{3/2}-\left(s_{i}-s_{i+1}\right)^{3/2}\right].
		\label{eq:si}
	\end{equation}
	Overdots denote time derivatives, and $s_{i}$ is the displacement of bead $i$ from its equilibrium position.
	
\begin{figure}[htbp]
		\centering
		\includegraphics[width=0.8\textwidth,trim=20 190 60 150,clip]{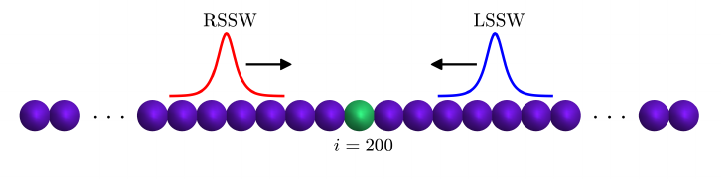}
		\caption{Schematic view of the simulation model. The collision center is located at bead $200$, which is green.}
		\label{fig:Fig1Model}
	\end{figure}

	In a simulation time step, the position and velocity of each bead are updated by integrating Eq. \eqref{eq:si}. The material of bead is stainless steel, and its parameters are as follows: Young's modulus $E=200~{\rm GPa}$, Poisson's ratio $\sigma=0.28$, density $\rho=7.9\times10^{3}~{\rm kg/{m^{3}}}$, radius $R=2.5~{\rm mm}$. The number of beads is $N=400$. To produce a SW, the leftmost bead of GC is set to an impulse velocity $v_{0}$ and the others are zero \cite{Sen1999PhysA268.644,Manciu1999PhysA274.588,Chong2017JPConMat29.413003}. The incident impulse spreads in GC after a certain traveling distance, composed of a leading right$-$traveling solitary wave (RSW) and a train of following secondary waves. This study examines the collision of HSWs by symmetrically placing two identical SWs at the sides of bead $200$, whose centers are located at bead $101$ and bead $299$, respectively. They are named as right$-$traveling scattered solitary wave (RSSW) and left$-$traveling scattered solitary wave (LSSW), as shown in Fig. \ref{fig:Fig1Model}. The collision center of HSWs is located at bead $200$ because of the identical symmetry of RSSW and LSSW. The study introduces $t_{i}$ as arrival time when the contact force $F_{i}$ between bead $i$ and bead $i+1$ gets to the maximum, which is normalized by $\left(k/m\right)^{-1/2}\left(2R\right)^{-1/4}$.
	
	\section{Theoretical analysis}
	\label{TheoAna}
	
	In Nesterenko's considerations, a theoretical solution of SW is suggested using long$-$wavelength approximation \cite{Nesterenko1983JAMTP24.733}. Keeping terms up to the fourth$-$order in spatial derivatives for Eq. \eqref{eq:si}, a continuous equation is derived using the strain $\xi=-s_{x}$ \cite{Nesterenko2001}.
	\begin{equation}
		\begin{aligned}
			&\xi_{tt}=c^{2}\left\{\xi^{3/2}+\frac{R^{2}}{3}\left[\left(\xi^{3/2}\right)_{xx}-\frac{3}{8}\left(\xi^{-1/2}\right)\xi_{x}^{2}\right]\right\}_{xx},\\
			&\xi>0, \qquad c^{2}=\frac{2E}{\pi\rho(1-\sigma^2)}=\frac{k}{m}\left(2R\right)^{5/2},
		\end{aligned}
		\label{eq:xitt}
	\end{equation}
	The constant $c$ is of the same order of magnitude as the bulk sound speed in grain. A single hump of periodic solution of Eq. \eqref{eq:xitt} with a fixed wavelength of five$-$bead diameters describes the proposed SW solution,
	\begin{equation}
		\xi=\left(\frac{5V_{\rm s}^{2}}{4c^{2}}\right)^2{\rm cos}^{4}\left(\frac{x-V_{\rm s}t}{\sqrt{10}R}\right),
		\label{eq:xi solution}
	\end{equation}
	where $x$ is the laboratory coordinate and $V_{\rm s}$ is the SW speed with a nonlinear dependence on the maximum bead velocity $v_{\rm m}$ \cite{Nesterenko2001,Nesterenko1983JAMTP24.733,Daraio2005PRE72.016603},
	\begin{equation}
		V_{\rm s}=\left(\frac{16}{25}\right)^{1/5}c^{4/5}v_{\rm m}^{1/5}=2R\left(\frac{4k}{5m}\right)^{2/5}v_{\rm m}^{1/5}.
		\label{eq:Vs vm}
	\end{equation}
	
	Considering the spatial compactness of a single SW, the quasi$-$particle model with $1.4m$ in mass and $1.4v_{\rm m}$ in velocity has constantly been introduced to describe its dynamics \cite{Nesterenko2001,Nesterenko1994JPhys4.C8729,Nesterenko1995JAMTP36.166,Daraio2004AIP706.197,Job2007GM10.13}. For the collision of two identical HSWs, one SW is regarded as static and approached by the other at the velocity of twice $1.4v_{\rm m}$. According to elastic theory, the duration time of the nonlinear collision of two identical HSWs is determined \cite{Landau1959},
	\begin{equation}
		\Delta t_{\rm G}=0.4\left(\frac{5\sqrt{2}\pi\rho}{4}\frac{1-\sigma^{2}}{E}\right)^{2/5}\frac{R}{v_{\rm m}^{1/5}}=0.4\left(\frac{5m}{8k}\right)^{2/5}v_{\rm m}^{-1/5}.
		\label{eq:elastic theory Delta t}
	\end{equation}
	
	For comparison purposes, we briefly revisit the nonlinear collision characteristics of HSWs as described by classical KdV equation in shallow water, which has been analytically solved \cite{Byatt-Smith1971JFM49.625,Whitham1974},
	\begin{equation}
		\begin{aligned}
			&\pm\eta_{t}+\left(1+\frac{3}{2}\alpha\eta\right)\eta_{x}+\frac{\beta^{2}}{6}\eta_{xxx}=0,\\
			&\alpha=\frac{A_{0}}{h}, \qquad \beta=k_{0}h,
		\end{aligned}
		\label{eq:KdV equation}
	\end{equation}
	where $x$, $t$, and $\eta$ are dimensionless horizontal coordinate, time, and free surface displacement. $A_{0}$ and $k_{0}$ are characteristic wave amplitude and wave number, and $h$ is water depth. Eq. \eqref{eq:KdV equation} has an exact analytical solution in SW form,
	\begin{equation}
		\eta={\rm{sech}}^{2}\left\{\frac{\sqrt{3\alpha}}{2\beta}\left[x\mp\left(1+\frac{\alpha}{2}\right)t\right]\right\},
		\label{eq:KdV solution}
	\end{equation}
	where the upper and lower signs denote the right and left traveling SWs in Eq. \eqref{eq:KdV equation} and Eq. \eqref{eq:KdV solution}.
	
	The reflection of a single SW at a vertical wall is equivalent to the nonlinear collision of two identical HSWs in fluid \cite{Maxworthy1976JFM76.177,Fenton1982JFM118.411,Power1984WM6.183,Byatt-Smith1988JFM197.503,Cooker1997JFM342.141,Chen2015EJMBF49.20}. Similarly, such considerations are also used in the studies for the collision of HSWs in GC \cite{Manciu2000PRE63.016614,Manciu2002PRE66.016616}. Compared with traveling a single SW, the collision of two HSWs in fluid leads to a space$-$phase shift. When two pieces of HSWs depart far away from each other, the uniform space$-$phase shift is theoretically derived using third$-$order approximation \cite{Su1980JFM98.509},
	\begin{subequations}
		\begin{align}
			&\Delta x_{\rm Un,R}=\left(\frac{A_{\rm L}}{3}\right)^{1/2}\left(1+\frac{1}{8}{A_{\rm L}}+\frac{3}{4}{A_{\rm R}}\right),\label{eq:x Un R}\\
			&\Delta x_{\rm Un,L}=-\left(\frac{A_{\rm R}}{3}\right)^{1/2}\left(1+\frac{1}{8}{A_{\rm R}}+\frac{3}{4}{A_{\rm L}}\right).\label{eq:x Un L}
		\end{align}
		\label{eq:UnPhaseShift}
	\end{subequations}
	At the position of the collision center, the phase shift in space occurs, defined as immediate space$-$phase shift,
	\begin{subequations}
		\begin{align}
			&\Delta x_{\rm Im,R}=\left(\frac{A_{\rm L}}{3}\right)^{1/2}\left(1+\frac{1}{8}{A_{\rm L}}+\frac{23}{4}{A_{\rm R}}\right),\label{eq:x Im R}\\
			&\Delta x_{\rm Im,L}=-\left(\frac{A_{\rm R}}{3}\right)^{1/2}\left(1+\frac{1}{8}{A_{\rm R}}+\frac{23}{4}{A_{\rm L}}\right).\label{eq:x Im L}
		\end{align}
		\label{eq:ImPhaseShift}
	\end{subequations}
	The space$-$phase shift $\Delta x$ and wave amplitude $A$ are reduced by the water depth $h$. The subscripts $\rm R$ and $\rm L$ denote the right and left traveling SWs, respectively.
	
	For the collision of single SW with static wall, two characteristic times, that is, attachment time $\Delta t_{\rm a}=t_{\rm a}-t_{\rm 0}$ and detachment time $\Delta t_{\rm d}=t_{\rm d}-t_{\rm 0}$, are two critical physical quantities describing the collision dynamics. $t_{\rm a}$ and $t_{\rm d}$ are when the incident wave crest arrives and leaves the wall, respectively. Under perfect reflection, the reflection is completed instantaneously, meaning that $t_{\rm a}$ and $t_{\rm d}$ collapse together. This moment is the reference time point $t_{0}$ \cite{Chen2015EJMBF49.20,Chambarel2009NonPG16.111}. The characteristic times can be written as \cite{Cooker1997JFM342.141}:
	\begin{subequations}
		\begin{align}
			&\Delta t_{\rm a}=t_{\rm a}-t_{0}=\frac{2}{\sqrt{3}}\left(-\kappa A^{-1/2}+\frac{1}{8}\left(2-\kappa\right)A^{1/2}\right),\label{eq:ta}\\
			&\Delta t_{\rm d}=t_{\rm d}-t_{0}=\frac{2}{\sqrt{3}}\left(\kappa A^{-1/2}+\frac{1}{8}\left(2+\kappa\right)A^{1/2}\right).\label{eq:td}
		\end{align}
		\label{eq:tatd}
	\end{subequations}
	Using the attachment time $\Delta t_{\rm a}$ and detachment time $\Delta t_{\rm d}$, the residence time $\Delta t_{\rm r}=t_{\rm d}-t_{\rm a}$ is introduced to describe the duration of collision,
	\begin{equation}
		\Delta t_{\rm r}=t_{\rm d}-t_{\rm a} =\frac{4\kappa}{\sqrt{3}}\left(A^{-1/2}+\frac{1}{8}A^{1/2}\right),
		\label{eq:tr}
	\end{equation}
	where $\kappa={\rm tanh}^{-1}\left(1/\sqrt{3}\right)$. The turning time $t'_{0}$ is also directly derived, which indicates the change time from the incident wave to the reflected wave. For the collision of two identical HSWs, maximum collision force appears at the time of $t'_{0}$. The time difference of $t'_{0}-t_{0}$ is the immediate time$-$phase shift at the position of the collision center of HSWs.
	\begin{equation}
		\Delta t_{\rm Im}=t'_{0}-t_{0}=\frac{1}{2\sqrt{3}}\left(A^{1/2}+\frac{43}{8}A^{3/2}\right).
		\label{eq:t0}
	\end{equation}
	The time is normalized by $\sqrt{h/g}$, and $g$ is the acceleration due to gravity.
	
	\section{Simulation results and discussion}
	\label{SimuResDis}
	
	\subsection{Dynamics of head$-$on collision}
	\label{DynaHeadOncoll}
	
\begin{figure}[htbp]
		\centering
		\includegraphics[width=0.8\textwidth]{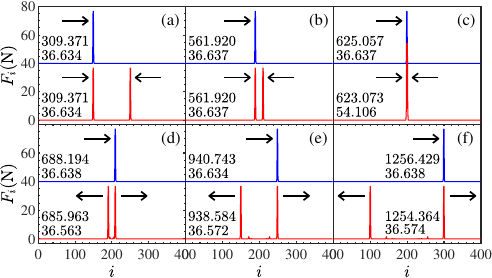}
		\caption{(color online) Collision force as a function of bead number for RSW (blue line) and RSSW, LSSW (red line) with impact velocity $v_{0}=1~{\rm m/s}$. (a)$-$(f) show the snapshots when the maximum forces on beads $150$, $190$, $200$, $210$, $250$, and $300$, and the corresponding arrival times are recorded. The data of RSW are shifted by 40$~\rm N$. The arrows stand for the traveling direction of SWs.}
		\label{fig:Fig2Force}
	\end{figure}

	The time evolution of a single RSW and two pieces of HSWs are plotted in Fig. \ref{fig:Fig2Force} when the maximum collision forces reach beads $150$, $190$, $200$, $210$, $250$, and $300$ to understand the collision dynamics of HSWs in GC. In each simulation run, the amplitude of the maximum force of RSW fluctuates slightly. The ratio of the instantaneous total energy to the initial total energy is also monitored, which fluctuates at the order of magnitude of $10^{-10}$.
	
	In Fig. \ref{fig:Fig2Force}, the maximum force and arrival time of RSSW and LSSW are completely symmetrical about the central bead of $200$. Thus, the following comparisons are only performed on the results of RSW and RSSW. The simulation results indicate three traveling stages of the two pieces of HSWs, namely, pre$-$in$-$phase traveling stage (pre$-$RSSW/pre$-$LSSW), central$-$collision stage, and post$-$in$-$phase traveling stage (post$-$RSSW/post$-$LSSW). At the pre$-$in$-$phase traveling stage, the RSSW travels freely in the GC because the collision has not occurred yet. The maximum force and arrival time of the RSSW are equal to those of the RSW shown in Fig. \ref{fig:Fig2Force} (a)(b). When the central$-$collision stage occurs, a clear nonlinear superposition of two pieces of HSWs is observed in Fig. \ref{fig:Fig2Force} (c), in which the collision reaches its central point, bead $200$. The collision between the HSWs yields a maximum force of $54.106~{\rm N}$, which is less than twice that of RSW. Furthermore, the collision decreases arrival time for the RSSW ($623.073$), as contrasted with the RSW ($625.057$). The leading$-$time phase equals $-1.984$, which is just the immediate time$-$phase shift. When the centers of two pieces of HSWs pass through each other, the arrival time of the RSSW is smaller than that of the RSW, as shown in Fig. \ref{fig:Fig2Force} (d)(e)(f). The former are $685.963$, $938.584$, and $1254.364$ for beads $210$, $250$, and $300$, respectively. The corresponding later are $688.194$, $940.743$, and $1256.429$, respectively. With increased separation distance between the centers of RSSW and LSSW, the two pieces of the HSWs loss mutual interaction completely and run into the stable post$-$in$-$phase traveling stage. In this stage, the time$-$phase difference between the RSSW and the RSW becomes a constant value, as shown in Fig. \ref{fig:Fig2Force} (e)(f).
	\begin{figure}[htbp]
		\centering
		\includegraphics[width=0.55\textwidth]{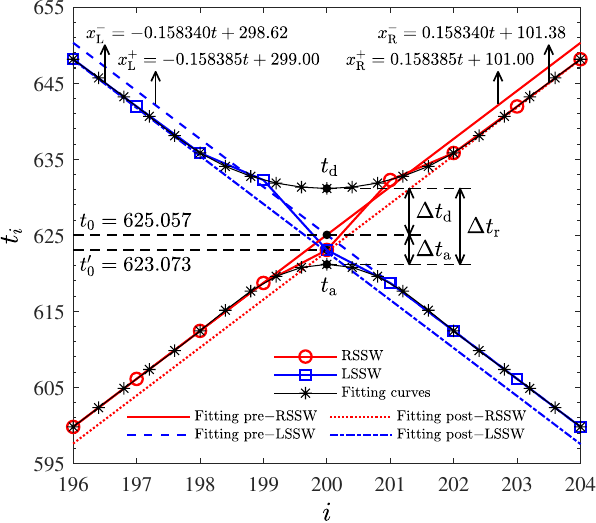}
		\caption{The arrival time as a function of bead number near bead $200$. Same parameters are used as those in Fig. \ref{fig:Fig2Force}. The meanings of lines are indicated in the legend. $t_{0}$ and $t'_{0}$ denote the simulation arrival times of bead 200 for RSW and HSWs, respectively. $\Delta t_{\rm a}$ and $\Delta t_{\rm d}$ represent the attachment and detachment times, respectively.}
		\label{fig:Fig3TimeAN}
	\end{figure}
	
	Based on the discussions of three traveling stages, the final leading$-$time phase is generated in the central$-$collision stage. To explore the collision dynamics of this stage, the arrival time is plotted as a function of bead number around bead $200$, as depicted in Fig. \ref{fig:Fig3TimeAN}. The arrival times of RSW and RSSW for beads $198$ to $207$ are given in Table 1. The absolute value of $\Delta t_{i}$ becomes larger from beads $202$ to $207$, but the absolute value of $\Delta t_{i+1}-\Delta t_{i}$ decreases. The difference of $\Delta t_{207}$ and $\Delta t_{206}$ is less than $10^{-2}$, which implies the central$-$collision stage is finished and the stable post$-$in$-$phase traveling stage starts. As shown in Table 1, the central$-$collision stage has an eight$-$bead width, namely, $199 \le i \le 206$, for RSSW. Then, the simulation results are fitted using the least$-$square method, $x_{\rm j}^{+}=a_{\rm j}^{+}t+b_{\rm j}^{+}$ for the pre$-$in$-$phase traveling stage ($i < 199$ for pre$-$RSSW or $i > 201$ for pre$-$LSSW) and $x_{\rm j}^{-}=a_{\rm j}^{-}t+b_{\rm j}^{-}$ for the post$-$in$-$phase traveling stage ($i > 206$ for post$-$RSSW or $i < 194$ for post$-$LSSW), respectively, where the subscripts $\rm j=\rm R,L$ represent the traveling direction of SWs. The fitting functions are $x_{\rm R}^{+}=0.158385t+101.00$ and $x_{\rm R}^{-}=0.158340t+101.38$ for pre$-$RSSW and post$-$RSSW, respectively, in which the fitted propagation velocities agree well with those of Eq. \eqref{eq:Vs vm} ($0.159631$ for pre$-$RSSW and $0.159584$ for post$-$RSSW). The propagation velocity of pre$-$RSSW is expected to be equal to that of RSW, whereas the post$-$RSSW's propagation velocity decreases owing to the collision of HSWs. The initial$-$space phase of $x_{\rm R}^{+}$ is $101.00$ consistent with the beginning location of the RSSW. The initial$-$space phase of post$-$RSSW, $101.38$, is obtained from the fitting result, and the leading$-$space phase is $0.38$. Similarly, the same fitting method is used for the LSSW, leading to symmetrical results of $x_{\rm L}^{+}=-0.158385t+299.00$ and $x_{\rm L}^{-}=-0.158340t+298.62$ for pre$-$LSSW and post$-$LSSW, respectively. The collision of HSWs causes a decrease in the propagation velocity of LSSW. However, the initial$-$space phase, $298.62$, is also obtained for post$-$LSSW, smaller than the initial$-$space phase of pre$-$LSSW, $299.00$. The corresponding leading$-$space phase is $0.38$, consistent with the result of RSSW.
	\begin{figure}[htbp]
		\centering
		\includegraphics[width=0.7\textwidth,trim=170 510 170 125,clip]{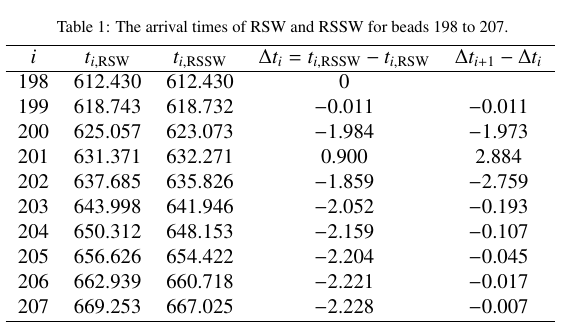} 
		\label{fig:Table1}
	\end{figure}
	
	In Fig. \ref{fig:Fig3TimeAN}, an asymmetrical characteristic of the central$-$collision stage is exhibited concerning bead $200$ for HSWs. The simulation arrival times of RSSW for beads $199$ and $200$ are less than those of RSW, which means the occurrence of the leading$-$time phase. Conversely, the arrival time of RSSW for bead $201$ is unexpectedly extended compared with that of RSW, which means a lagging$-$time phase. For bead $202$ and the latter beads of RSSW, a leading$-$time phase reappears, in which the arrival time of RSSW is less than that of RSW. Thus, the central$-$collision stage can be divided into two sub$-$stages (i.e., incident collision process and separation collision process). The former exhibits a compression process and an expansion process in the latter. The leading$-$time phase covers the entire compression process and the latter half of the expansion process, and the lagging$-$time phase only appears at the beginning half of the expansion process. This collision characteristic is reversed to our previous observations for the collision of HSWs in a compressed GC \cite{Wu2020CPL37.074501}. However, the leading$-$time phase is more prominent than the lagging$-$time phase. A leading phase shift of time occurs for the HSWs after the collision, consistent with that in a compressed GC.
	
	\subsection{Space$-$phase shift}
	\label{SpaPhaShi}
	
	From the observations presented above, the nonlinear collision of HSWs leads to leading phase shifts in time and space though the amplitude of the propagation velocity of SWs decreases after the collision. To compare the space$-$phase shift of HSWs in nonintegrable GC and integrable fluid, we employ the same analysis method used in fluid to obtain fitting functions for pre$-$in$-$phase and post$-$in$-$phase traveling stages \cite{Craig2006POF18.057106}. The immediate space$-$phase shift at the collision center is calculated by the difference in the intercepts of the fitting lines at the arrival time of maximum collision force $t'_{0}$,
	\begin{equation}
		\Delta x_{\rm Im,j}=\left(a_{\rm j}^{-}-a_{\rm j}^{+}\right)t'_{0}+\left(b_{\rm j}^{-}-b_{\rm j}^{+}\right), \quad \rm j=\rm R,L.
		\label{eq:Delta x Im,j}
	\end{equation}
	The fitting results of RSSW and LSSW in Fig. \ref{fig:Fig3TimeAN} give the same immediate space$-$phase shifts (i.e., $\Delta x_{\rm Im,R}=0.3562$ and $\Delta x_{\rm Im,L}=-0.3562$). 
	\begin{figure}[htbp]
		\centering
		\includegraphics[width=0.5\textwidth]{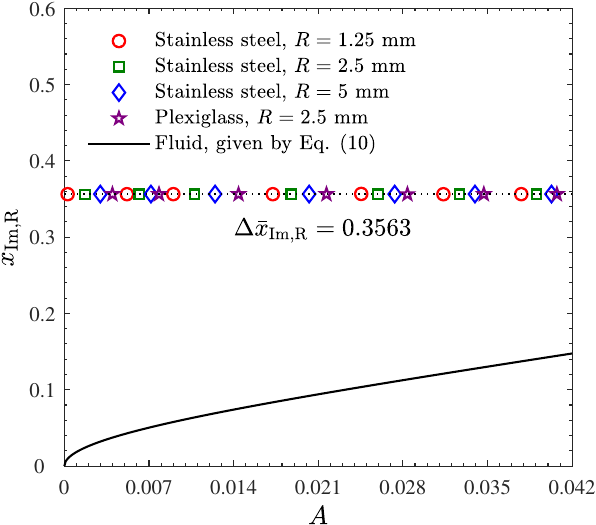}
		\caption{Immediate space$-$phase shift of RSSW as a function of the wave amplitude of HSWs. Different beads are indicated in the legend. Black solid line is for the immediate space$-$phase shift of fluid, given by Eq. \eqref{eq:ImPhaseShift}.}
		\label{fig:Fig4ImPhaseShift}
	\end{figure}	
	
	We also conducted a series of similar simulations on the collisions of HSWs with different amplitudes normalized by $2R$. The immediate space$-$phase shift of RSSW against the wave amplitude of HSWs is plotted in Fig. \ref{fig:Fig4ImPhaseShift}. Unexpectedly, the immediate space$-$phase shift in GC remains constant $0.3563$, whereas a monotonic increase is found in fluid. This unique collision characteristic of HSWs in GC can be reasonably explained using the quasi$-$particle arguments in Section \ref{TheoAna}. The immediate space$-$phase shift is the product of the propagation velocity, given by Eq. \eqref{eq:Vs vm}, and the collision duration, given by Eq. \eqref{eq:elastic theory Delta t}. The theoretical result is $0.3031$, which does not depend on the wave amplitude of HSWs. Moreover, three stainless steel beads with $R=1.25,2.5$ and $5~{\rm mm}$ and one plexiglass bead with $R=2.5~{\rm mm}$ are used in the simulations. All results fall on the same straight line, with an average of $ \Delta {\bar x}_{\rm Im,R}=0.3563$. The parameters of the plexiglass bead are as follows: Young's modulus $E=71.7~{\rm GPa}$, Poisson's ratio $\sigma=0.23$, density $\rho=2.5\times10^{3}~{\rm kg/{m^{3}}}$ and radius $R=2.5~{\rm mm}$ \cite{Du2020IJP98.1249}.
	\begin{figure}[htbp]
		\centering
		\includegraphics[width=0.5\textwidth]{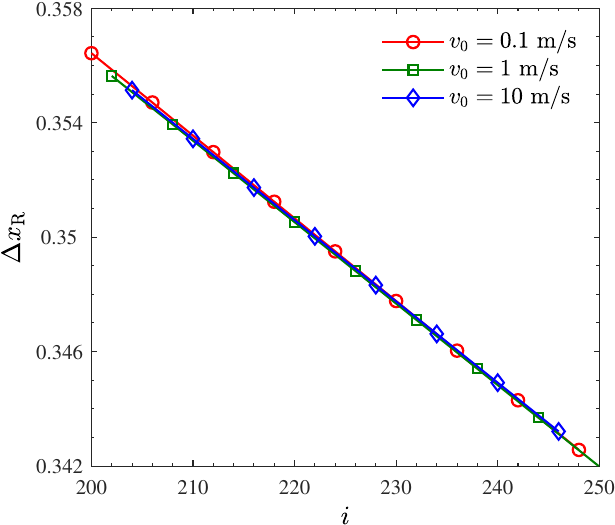}
		\caption{Space$-$phase shift of RSSW as a function of bead location. The stainless steel bead radius is $2.5~{\rm mm}$. The red circles, green squares, and blue diamonds are for impact velocities of $0.1~{\rm m/s}$, $1~{\rm m/s}$, and $10~{\rm m/s}$, respectively.}
		\label{fig:Fig5RSSWPhaseShift}
	\end{figure}
	
	Two determined factors for the space$-$phase shift after collision are the immediate space$-$phase shift at the collision center and the propagation velocity difference between RSW and post$-$RSSW. When the collision of HSWs occurs in fluid, the SWs can recover their original incident waveform after the collision \cite{Su1980JFM98.509,Fenton1982JFM118.411,Chen2014JFM749.577,Tong2019WM88.34}. The uniform space$-$phase shift given by Eq. \eqref{eq:UnPhaseShift} is only determined by the wave amplitude of HSWs, independent of the measurement position. However, for the case of GC, as observed in Fig. \ref{fig:Fig2Force}, the nonlinear scattering effect in the central$-$collision stage leads to a decrease in the amplitude of post$-$RSSW compared with that of RSW. In Fig. \ref{fig:Fig4ImPhaseShift}, the results show that the wave amplitude of HSWs does not influence the immediate space$-$phase shift. On the other hand, the post$-$RSSW travels stably at a lower propagation velocity in GC after the collision. Thus, the space$-$phase shift after the collision is expected to decrease linearly with an increase in the traveling distance of the post$-$RSSW, as confirmed by the simulation results shown in Fig. \ref{fig:Fig5RSSWPhaseShift} for impact velocities of $v_{0}=0.1,1$ and $10~{\rm m/s}$.
	
	\subsection{Collision characteristic time of HSWs}
	\label{CollCharaTimeHSWs}
	
	The above observations have demonstrated that the nonlinear scattering effect occurs in the central$-$collision stage for GC and fluid. The space$-$phase shift of the former is independent of wave amplitude whereas that of the latter increases with the wave amplitude of HSWs. For complete comparison purposes, we next compare the characteristic time of the central$-$collision stage for GC and fluid. The attachment and detachment times are introduced for GC using the similar analysis method as that in fluid, i.e., Eq. \eqref{eq:tatd}. $t_{\rm a}$ and $t_{\rm d}$ for GC are obtained by the third$-$order spline difference method shown in Fig. \ref{fig:Fig3TimeAN}.
	
\begin{figure}[htbp]
		\centering
		\includegraphics[width=0.75\textwidth]{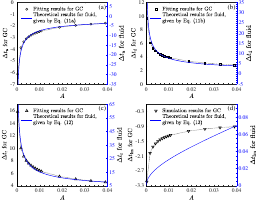}
		\caption{Collision characteristic time as a function of the wave amplitude of HSWs. The stainless steel bead radius is $2.5~{\rm mm}$. (a) attachment time $\Delta t_{\rm a}$; (b) detachment time $\Delta t_{\rm d}$; (c) residence time $\Delta t_{\rm r}$; (d) immediate time$-$phase shift $\Delta t_{\rm Im}$. The black circles, squares, upward triangles, and downward triangles are for the results of GC, respectively. The blue solid lines are for the theoretical results of fluid. The dotted lines are guides for eyes.}
		\label{fig:Fig6CharaTime}
	\end{figure}

	Theoretical results in fluid from Eq. \eqref{eq:tatd} to Eq. \eqref{eq:tr}, as well as the fitting results in GC are shown in Fig. \ref{fig:Fig6CharaTime} (a)(b)(c). For the collision of HSWs in GC, the collision center is located at bead $200$ due to the symmetry of HSWs. Bead $200$ stays stationary in the entire collision process and can be regarded as an infinite$-$mass bead at the center of GC \cite{Manciu2000PRE63.016614, Manciu2002PRE66.016616}. Similar to the fluid analysis, the arrival time of bead $200$ for RSW is chosen as the reference time point $t_{0}$, which is the turning point between the incident and reflection waves in perfect reflection.
	
	In Fig. \ref{fig:Fig6CharaTime} (a), the attachment time $\Delta t_{\rm a}$ is plotted as a function of the wave amplitude of HSWs. Similar dependence is observed for the fitting results in GC and the theoretical results in fluid predicted by Eq. \eqref{eq:ta}, both of which are negative values. Compared with perfect reflection, RSSW has a larger propagation velocity, resulting in an earlier arrival time at bead $200$. Fig. \ref{fig:Fig6CharaTime} (b) plots the dependence of the detachment time $\Delta t_{\rm d}$ on the wave amplitude of HSWs. Similar results are observed for GC and fluid, which are positive values. RSSW has a larger detachment time compared with the perfect reflection. In Fig. \ref{fig:Fig6CharaTime} (c), the dependence of residence time $\Delta t_{\rm r}$ on the wave amplitude of HSWs is plotted, which is the time difference between $t_{\rm d}$ and $t_{\rm a}$. The larger amplitude of HSWs decreases monotonically the residence time in GC and fluid.
	
	In Fig. \ref{fig:Fig6CharaTime} (d), we plot the dependence of the immediate time$-$phase shift $\Delta t_{\rm Im}$ on the wave amplitude of HSWs for GC and fluid. The reference time point is also placed at the turning point of perfect reflection, as those in Fig. \ref{fig:Fig6CharaTime} (a) and (b). The negative and positive values of $\Delta t_{\rm Im}$ denote that the collision of HSWs leads to a leading$-$and lagging$-$time phase, respectively. The smaller absolute value of $\Delta t_{\rm Im}$ means that the collision of HSWs approaches the perfect reflection. As shown in Fig. \ref{fig:Fig6CharaTime} (d), $\Delta t_{\rm Im}$ increases with an increase in the wave amplitude of HSWs for GC and fluid. However, an unexpected opposite result is observed, i.e., $\Delta t_{\rm Im}$ for GC is negative, whereas $\Delta t_{\rm Im}$ for fluid is positive. The collision of HSWs in GC approaches the perfect reflection as the wave amplitude of HSWs increases while that for fluid is reversed. This opposite result may originate from the difference in research models, the GC model is a compressible Hertz system, whereas the fluid is assumed to be incompressible in previous studies \cite{Su1980JFM98.509,Fenton1982JFM118.411,Power1984WM6.183,Byatt-Smith1988JFM197.503,Cooker1997JFM342.141,Chen2014JFM749.577,Tong2019WM88.34,Chambarel2009NonPG16.111,Chen2015EJMBF49.20}.
	%\newpage
	\section{Conclusions}
	\label{Conc}
	
	In this study, the nonlinear interaction of two identical head$-$on solitary waves is numerically simulated in a sonic$-$vacuum granular chain, which is nonintegrable. The phenomenological solution of solitary wave proposed by Nesterenko is revisited and the quasi$-$particle model is used to describe the collision of solitary waves. We also revisit the theoretical analysis of the classical Korteweg$-$de Vries equation in integrable fluid (i.e., phase shifts in space and time). Similar analyses are performed on the simulation results of granular chain, and the major findings are summarized as follows.
	
	\begin{itemize}
		\item[$\bullet$] For the collision of head$-$on solitary waves in a granular chain, three stages are identified to describe the interaction processes: pre$-$in$-$phase traveling stage, central$-$collision stage, and post$-$in$-$phase traveling stage. The solitary waves can travel freely in the pre$-$and post$-$in$-$phase traveling stages. The nonlinear scattering effect appears in the central$-$collision stage, leading to a decrease in the amplitude of solitary waves.
		
		\item[$\bullet$] The central$-$collision stage has an asymmetrical collision characteristic, i.e., incident and separation collision processes, which correspond to the compression and expansion processes of solitary waves. In the entire compression process and the latter half of the expansion process, the leading$-$time phase occurs, and the lagging$-$time phase happens at the beginning half of the expansion process. The accumulation result leads to leading phase shifts in time and space after the collision of solitary waves.
		
		\item[$\bullet$] In the granular chain, the immediate space$-$phase shift is found independent of wave amplitude and material parameters, which can be reasonably explained by the quasi$-$particle model. The space$-$phase shift after the collision is only determined by the measurement position rather than the wave amplitude. The corresponding results are reversed in fluid. In the former, the nonlinear scattering effect results in a decrease in the amplitude of solitary waves. The solitary waves recover their original incident waveform after the collision in the latter.
		
		\item[$\bullet$] The attachment, detachment, and residence times decrease with the increasing amplitude of solitary waves for granular chain and fluid. However, an opposite result occurs for immediate time$-$phase shift, i.e., a leading shift for granular chain and a lagging shift for fluid.
	\end{itemize}
	
	The present work lends insights into the nature of solitary waves interactions for nonintegrable granular chain by comparing the results with those of integrable fluid. This treatment provides a powerful analysis method to measure a variety of mechanical waves in integrable and nonintegrable systems accurately. The unique interaction properties of solitary waves in a granular chain stem from its special interaction force. These innovative observations pave the way to design energy$-$harvest and shock$-$protection devices.
	
	\section*{Acknowledgements}
	This work are financially supported by the National Natural Science Foundation of China (Grant No. 11574153) and the foundation of the Ministry of Industry and Information Technology of China (Grant No. TSXK2022D007).


\noindent
	\begin{thebibliography}{99}
		%1-5
		\bibitem{Osborne1980SCI208.451}A. R. Osborne, and T. L. Burch, Internal Solitons in the Andaman Sea. Science {\bf 208}, 451$-$460(1980).
		
		\bibitem{Shukla1984POF27.327}P. K. Shukla, M. Y. Yu, and N. L. Tsintsadze, Intense solitary laser pulse propagation in a plasma. Phys. Fluids {\bf 27}, 327(1984).
		
		\bibitem{Shukla2007PLA367.120}N. Shukla, and P. K. Shukla, A new purely growing instability in a strongly magnetized nonuniform pair plasma. Phys. Lett. A {\bf 367}, 120$-$122(2007).
		
		\bibitem{Schwarz1999PRL83.223}U. T. Schwarz, L. Q. English, and A. J. Sievers, Experimental Generation and Observation of Intrinsic Localized Spin Wave Modes in an Antiferromagnet. Phys. Rev. Lett. {\bf 83}, 223$-$226(1999).
		
		\bibitem{Xie2000PRL84.5435}A. H. Xie, L. van der Meer, W. Hoff, and R. H. Austin, Long$-$Lived Amide I Vibrational Modes in Myoglobin. Phys. Rev. Lett. {\bf 84}, 5435$-$5438(2000).
		%6-10
		\bibitem{Fleischer2003Nature422.147}J. W. Fleischer, M. Segev, N. K. Efremidis, and D. N. Christodoulides, Observation of two$-$dimensional discrete solitons in optically$-$induced nonlinear photonic lattices. Nature {\bf 422}, 147$-$150(2003).
		
		\bibitem{Wazwaz2009}A. W. WazWaz, \emph{Partial Differential Equations and Solitary Waves Theory}. Higher Education Press, 2009.
		
		\bibitem{Sen2008PR462.21}S. Sen, J. Hong, J. Bang, E. Avalos, and R. Doney, Solitary waves in the granular chain. Phys. Rep. {\bf 462}, 21$-$66(2008).
		
		\bibitem{Rosas2018PhysRep753.1}A. Rosas, and K. Lindenberg, Pulse propagation in granular chains. Phys. Rep. {\bf 735}, 1$-$37(2018).
		
		\bibitem{Korteweg1895PM39.422}D. J. Korteweg, and G. de Vries, On the change of form of long waves advancing in a rectangular canal, and on a new type of long stationary waves. Philos. Mag. {\bf 91}, 1007$-$1028(2011).
		%11-15
		\bibitem{Nesterenko1983JAMTP24.733}V. F. Nesterenko, Propagation of nonlinear compression pulses in granular media. J. Appl. Mech. Tech. Phys. {\bf 24}, 733$-$743(1983).
		
		\bibitem{Nesterenko2018PTRSA376.20170130}V. F. Nesterenko, Waves in strongly nonlinear discrete systems. Phil. Trans. R. Soc. A {\bf 376}, 20170130(2018).
		
		\bibitem{Fermi1955}E. Fermi, J. Pasta, and S. Ulam, Studies of nonlinear problems. Tech. rep. Los Alamos Scientific Laboratory Report NO. LA$-$1940, 1955, reprinted in Lect. Appl. Math. {\bf 15}, 143$-$156(1974).
		
		\bibitem{Vainchtein2022PhysD434.133252}A. Vainchtein, Solitary waves in FPU$-$type lattices. Phys. D {\bf 434}, 133252(2022).
		
		\bibitem{Nesterenko2001}V. F. Nesterenko, \emph{Dynamics of Heterogeneous Materials}. Springer$-$Verlag, 2001.
		%16-20
		\bibitem{Sen1998PRE57.2386}S. Sen, M. Manciu, and J. D. Wright, Solitonlike pulses in perturbed and driven Hertzian chains and their possible applications in detecting buried impurities. Phys. Rev. E {\bf 57}, 2386$-$2397(1998).
		
		\bibitem{Sen1999PhysA268.644}S. Sen, and M. Manciu, Discrete Hertzian chains and solitons. Phys. A {\bf 268}, 644$-$649(1999).
		
		\bibitem{Manciu1999PhysA274.588}M. Manciu, S. Sen, and A. J. Hurd, The propagation and backscattering of soliton$-$like pulses in a chain of quartz beads and related problems. (I). Propagation. Phys. A {\bf 274}, 588$-$606(1999).
		
		\bibitem{Chong2017JPConMat29.413003}C. Chong, M. A. Porter, P. G. Kevrekidis, and C. Daraio, Nonlinear coherent structures in granular crystals. J. Phys.: Condens. Matter {\bf 29}, 413003(2017).
		
		\bibitem{Manciu2000PRE63.016614}M. Manciu, S. Sen, and A. J. Hurd, Crossing of identical solitary waves in a chain of elastic beads. Phys. Rev. E {\bf 63}, 016614(2000).
		%21-25
		\bibitem{Manciu2002PRE66.016616}F. S. Manciu, and S. Sen, Secondary solitary wave formation in systems with generalized Hertz interactions. Phys. Rev. E {\bf 66}, 016616(2002).
		
		\bibitem{Avalos2009PRE79.046607}E. Avalos, and S. Sen, How solitary waves collide in discrete granular alignments. Phys. Rev. E {\bf 79}, 046607(2009).
		
		\bibitem{Santibanez2011PRE84.026604}F. Santibanez, R. Munoz, A. Caussarieu, S. Job, and F. Melo, Experimental evidence of solitary wave interaction in Hertzian chains. Phys. Rev. E {\bf 84}, 026604(2011).
		
		\bibitem{Shen2014PRE90.022905}Y. Shen, P. G. Kevrekidis, S. Sen, and A. Hoffman, Characterizing traveling$-$wave collisions in granular chains starting from integrable limits: The case of the Korteweg$–$de Vries equation and the Toda lattice. Phys. Rev. E {\bf 90}, 022905(2014).
		
		\bibitem{Wang2018CPB27.044501}F. G. Wang, Y. Y. Yang, J. F. Han, and W. S. Duan, Head$-$on collision between two solitary waves in one$-$dimensional bead chain. Chin. Phys. B {\bf 27}, 044501(2018).
		%26-30
		\bibitem{Wu2020CPL37.074501}Q. Q. Wu, X. Y. Liu, T. F. Jiao, S. Sen, and D. C. Huang, Head$-$on Collision of Solitary Waves Described by the Toda Lattice Model in Granular Chain. Chin. Phys. Lett. {\bf 37}, 074501(2020).
		
		\bibitem{Zhang2021IJMS191.106073}W. Zhang, and J. Xu, Tunable traveling wave properties in one$-$dimensional chains composed from hollow cylinders: From compression to rarefaction waves. Int. J. Mech. Sci. {\bf 191}, 106073(2021).
		
		\bibitem{Tichler2013PRL111.048001}A. M. Tichler, L. R. Gomez, N. Upadhyaya, X. Campman, V. F. Nesterenko, and V. Vitelli, Transmission and Reflection of Strongly Nonlinear Solitary Waves at Granular Interfaces. Phys. Rev. Lett. {\bf 111}, 048001(2013).
		
		\bibitem{Liu2015PRE92.013202}S. W. Liu, Y. Y. Yang, W. S. Duan, and L. Yang, Pulse reflection and transmission due to impurities in a granular chain. Phys. Rev. E {\bf 92}, 013202(2015).
		
		\bibitem{Du2020IJP98.1249} W. Q. Du, Y. Y. Yang, J. F. Han, and W. S. Duan, Reflection and transmission of the incident wave due to impurities in the bead chain. Indian J. Phys. {\bf 94}, 1249$-$1253(2020).
		%31-35
		\bibitem{Yang2016AIP6.075317}Y. Y. Yang, S. W. Liu, Q. Yang, Z. B. Zhang, W. S. Duan, and L. Yang, Solitary waves propagation described by Korteweg$-$de Vries equation in the granular chain with initial prestress. AIP Advances {\bf 6}, 075317(2016).
		
		\bibitem{Su1980JFM98.509}C. H. Su, and R. M. Mirie, On head$-$on collisions between two solitary waves. J. Fluid Mech. {\bf 98}, 509$-$525(1980).
		
		\bibitem{Chen2014JFM749.577}Y. S. Chen, and H. Yeh, Laboratory experiments on counter$-$propagating collisions of solitary waves. Part 1. Wave interactions. J. Fluid Mech. {\bf 749}, 577$-$596(2014).
		
		\bibitem{Tong2019WM88.34}C. Tong, Y. L. Shao, F. C. W. Hanssen, Y. Li, B. Xie, and Z. L. Lin, Numerical analysis on the generation, propagation and interaction of solitary waves by a Harmonic Polynomial Cell Method. Wave Motion {\bf 88}, 34$-$56(2019).
		
		\bibitem{Chambarel2009NonPG16.111}J. Chambarel, C. Kharif, and J. Touboul, Head$-$on collision of two solitary waves and residual falling jet formation. Nonlin. Processes Geophys. {\bf 16}, 111$-$122(2009).
		%36-40
		\bibitem{Deng2020Chaos30.043101}G. Deng, G. Biondini, and S. Sen, Interactions of solitary waves in integrable and nonintegrable lattices. Chaos {\bf 30}, 043101(2020).
		
		\bibitem{Maxworthy1976JFM76.177}T. Maxworthy, Experiments on collisions between solitary waves. J. Fluid Mech. {\bf 76}, 177$-$185(1976).
		
		\bibitem{Fenton1982JFM118.411}J. D. Fenton, and M. M. Rienecker, A Fourier method for solving nonlinear water$-$water problems: application to solitary$-$wave interactions. J. Fluid Mech. {\bf 118}, 411$-$443(1982).
		
		\bibitem{Power1984WM6.183}H. Power, and A. T. Chwang, On reflection of a planar solitary wave at a vertical wall. Wave Motion {\bf 6}, 183$-$195(1984).
		
		\bibitem{Byatt-Smith1988JFM197.503}J. G. B. Byatt$-$Smith, The reflection of a solitary wave by a vertical wall. J. Fluid Mech. {\bf 197}, 503$-$521(1988).
		%41-45
		\bibitem{Cooker1997JFM342.141}M. J. Cooker, P. D. Weidman, and D. S. Bale, Reflection of a high$-$amplitude solitary wave at a vertical wall. J. Fluid Mech. {\bf 342}, 141$-$158(1997).
		
		\bibitem{Chen2015EJMBF49.20}Y. Y. Chen, C. Kharif, J. H. Yang, H. C. Hsu, J. Touboul, and J. Chambarel, An experimental study of steep solitary wave reflection at a vertical wall. Eur. J. Mech. B$-$Fluid. {\bf 49}, 20$-$28(2015).
		
		\bibitem{ZhangJ2014POP21.103706}J. Zhang, Y. Yang, Y. X. Xu, L. Yang, X. Qi, and W. S. Duan, The study of the Poincare$-$Lighthill$-$Kuo method by using the particle$-$in$-$cell simulation method in a dusty plasma. Phys. Plasmas {\bf 21}, 103706(2014).
		
		\bibitem{Qi2014POP21.082118}X. Qi, Y. X. Xu, W. S. Duan, L. Y. Zhang, and L. Yang, Particle$-$in$-$cell simulation of the head$-$on collision between two ion acoustic solitary waves in plasmas. Phys. Plasmas {\bf 21}, 082118(2014).
		
		\bibitem{ZhangJ2016CPL33.065202}J. Zhang, X. Qi, H. Zhang, and W. S. Duan, Particle$-$in$-$cell Simulation of the Reflection of a Korteweg$-$de Vries Solitary Wave and an Envelope Solitary Wave at a Solid Boundary. Chin. Phys. Lett. {\bf 33}, 065202(2016).
		%46-50
		\bibitem{Landau1959}L. D. Landau, and E. M. Lifshitz, \emph{Theory of Elasticity}, Pergamon Press, 1959.
		
		\bibitem{Daraio2005PRE72.016603}C. Daraio, V. F. Nesterenko, E. B. Herbold, and S. Jin, Strongly nonlinear waves in a chain of Teflon beads. Phys. Rev. E {\bf 72}, 016603(2005).
		
		\bibitem{Nesterenko1994JPhys4.C8729}V. F. Nesterenko, Solitary waves in discrete media with anomalous compressibility and similar to ``sonic vacuum''. J. Phys. IV {\bf 4}, C8$-$729$-$C8$-$734(1994).
		
		\bibitem{Nesterenko1995JAMTP36.166}V. F. Nesterenko, A. N. Lazaridi, and E. B. Sibiryakov, The decay of soliton at the contact of two ``acoustic vacuums''. J. Appl. Mech. Tech. Ph+ {\bf 36}, 166$-$168(1995).
		
		\bibitem{Daraio2004AIP706.197}C. Daraio, V. Nesterenko, and S. Jin, Strongly nonlinear waves in 3D phononic crystals. AIP Conf. Proc. {\bf 706}, 197$-$200(2004).
		%51-
		\bibitem{Job2007GM10.13}S. Job, F. Melo, A. Sokolow, and S. Sen, Solitary wave trains in granular chains: experiments, theory and simulations. Granular Matter {\bf 10}, 13$-$20(2007).
		
		\bibitem{Byatt-Smith1971JFM49.625}J. G. B. Byatt$-$Smith, An integral equation for unsteady surface waves and a comment on the Boussinesq equation. J. Fluid Mech. {\bf 49}, 625$-$633(1971).
		
		\bibitem{Whitham1974}G. B. Whitham, \emph{Linear and Nonlinear Waves}, Wiley, 1974.
		
		\bibitem{Craig2006POF18.057106}W. Craig, P. Guyenne, J. Hammack, D. Henderson, and C. Sulem, Solitary water wave interactions. Phys. Fluids {\bf 18}, 057106(2006).
		
	\end{thebibliography}
\end{document}